\documentclass{aa}
\usepackage{graphicx}
\usepackage{amssymb}
\usepackage{amsmath}
\newcommand{\harps}{{\textsc{harps}}}\newcommand{\corot}{{\textsc{CoRoT}}}
\newcommand{\cible}{{HD~203608}}

\newcommand{\zeroun}{\delta\nu_{01}}
\newcommand{\zerodeux}{\delta\nu_{02}}
\newcommand{\ind}[1]{_{\rm #1}}

\def\m2s2{\,m$^{2}$\,s$^{-2}$} %m2.s -2
\def\nunl{\nu_{n,\ell}}
\def\aov{\alpha_{\hbox{\rm\small ov}}}
\def\gradeps{\nabla\varepsilon}

\begin{document}
\title{Survival of a convective core in low-mass solar-like pulsator \cible}
%\subtitle{}
\titlerunning{Asteroseismic study of HD 203608}
\author{
S. Deheuvels\inst{1}\and
E. Michel\inst{1}\and
M.J. Goupil\inst{1}\and
J.P. Marques\inst{1}\and
B. Mosser\inst{1}\and
M.A. Dupret\inst{1}\and
Y. Lebreton\inst{1}\and
B. Pichon\inst{2}\and
P. Morel\inst{2}
}

\offprints{S. Deheuvels\\ \email{sebastien.deheuvels@obspm.fr}
}

\institute{LESIA, CNRS UMR 8109,  Observatoire de Paris, Universit\'e Paris 6, Universit\'e Paris 7, 92195 Meudon cedex, France\\
\and
Laboratoire Cassiop\'ee, CNRS UMR 6202, Observatoire de la C{\^o}te d'Azur,  BP 4229, 06304 Nice Cedex 4, France
}

\date{Submitted ...}%; accepted March 16, 1997}

\abstract{A 5-night asteroseismic observation of the F8V star \cible\ was conducted in August 2006 with \harps, followed by an analysis
of the data, and a preliminary modeling of the star (Mosser et al. 2008). The stellar parameters were significantly constrained, but the behavior of one of the seismic 
indexes (the small spacing $\zeroun$) could not be fitted with the observed one, even with the best considered models.}%
{We study the possibility of improving the agreement between models and observations by changing the physical properties of
the inner parts of the star (to which $\zeroun$ is sensitive).}%
{We show that, in spite of its low mass, it is possible to produce models of \cible\ with a convective core. No such model was considered in the
preliminary modeling. In practice, we obtain these models here by assuming some extra mixing at the edge
of the early convective core. We optimize the model parameters using the Levenberg-Marquardt algorithm.}%
{
The agreement between the new best model with a convective core and the observations is much better than for the models without. 
All the observational parameters are fitted within 1-$\sigma$ observational error bars. This is the first observational evidence of a convective core
in an old and low-mass star such as \cible. In standard models of low-mass stars, the core withdraws shortly after the ZAMS. The survival of the core until the present age of \cible\ provides
very strong constraints on the size of the mixed zone associated to the convective core. Using overshooting as a proxy to model the processes of transport at the
edge of the core, we find that to reproduce both global and seismic observations, we must have $\aov=0.17\pm0.03\,H_p$ for \cible.
We revisit the process of the extension of the core lifetime due to overshooting in the particular case of \cible.
 }%
{}

\keywords{Stars: evolution -- Stars: oscillations -- Stars: interiors }

\maketitle
%________________________________________________________________
\section{Introduction}

For main sequence stars massive enough to show a convective core ($M\gtrsim1.1M_{\odot}$ for solar-like metallicity), the associated mixed region plays the role of a reservoir
for nuclear reactions. The evolution pace of these stars and the time they spend on the main sequence depend directly on the size of this reservoir. The imprecise
knowledge we have of the mixing processes, particularly at the boundary of the core, generates large uncertainties on the extension of the mixed core and subsequently
on the stellar age and mass for a given set of surface parameters. Among the processes of transport of chemical elements that could contribute to the creation of a mixed zone
beyond the edge of the convective core, overshooting is the one invoked most often.

In the deep interior, convective elements rise adiabatically. They are accelerated until they reach the position of convective stability, i.e. $\nabla\ind{rad}=\nabla\ind{ad}$.
Then, the buoyancy forces cause a braking of the eddies in the radiative region. It is, however, unlikely that they should stop abruptly at the boundary between the
two regimes. They might penetrate, over a distance $d\ind{ov}$, in the regions of stability owing to their inertia, and generate a region of mixing beyond the edge
of the core. This phenomenon has been investigated by several authors (see \cite{1991A&A...252..179Z} and references therein), but no satisfying theoretical or
numerical description have been proposed. In practice, this region is modeled as an adiabatic layer above the core, whose thickness is a fraction $\aov$ of the pressure scale height $H_p$ 
($d\ind{ov}=\aov H_p$) and where the elements are mixed. While it is admitted that $\aov$ is only a crude account for the complex processes of mixing at the boundary of the
convective core, it is convenient and usual in the modeling of stellar interiors to adopt a representation of these processes depending on this parameter alone.
Different studies have led to a wide range of $\aov$: between zero (\cite{1986A&A...164...45L}) and about 2 (\cite{1985A&A...150..133X}). 
In fact, it is currently admitted that different values of $\aov$ might be needed to model stars of different masses and ages (see \textit{e.g.} \cite{2007A&A...475.1019C}).
We therefore do not have precise knowledge of the amount of mixing at the edge of the core, and
it is one of the main goals of asteroseismology to constrain it with observations (see \cite{1995IAUS..166..135L}, \cite{2006ESASP1306...39M}).

For intermediate-mass and high-mass stars, $d\ind{ov}$ is admitted to play an important role (\cite{1976A&A....47..389M}). The case of low-mass stars is not as clear. 
When they reach the ZAMS, these stars present a small convective core that disappears almost immediately. It has already been mentioned that an
extra mixing at the edge of this early convective core might increase its longevity, by providing first more $^{12}$C, and then more $^3$He in the center (Roxburgh 1985).
In the specific case of the Sun, core overshooting was added in the models, but it was concluded that it had no relevant impact on the Sun's present structure, 
unless we add an unreasonable amount of extra mixing. Later on, stellar models of low-mass stars suggested that the overshooting at the edge 
of the core could make it survive almost until the end of the main sequence, although the phenomenon was not explained (\cite{1993ASPC...40..454M}).

%This matter was first studied when trying to solve the problem of the unexpectedly low value
%of the observed neutrino flux emmited by the Sun. Several studies generated an \textit{ad hoc} mixed core in their solar models (\textit{e.g.} \cite{1971ApJ...165..171S}, 
%\cite{1990ApJ...349..641S}). Later on, other stellar models for low-mass stars suggested that the presence of overshooting at the edge of a convective core could significantly
%increase the lifetime of this core (see \cite{1993ASPC...40..454M}). However, we still require to understand the physical processes which
%extend the longevity of the core, and to confirm this phenomenon with observations.

In this article, we revisit this phenomenon in the case of \cible, a low-mass F8V star that presents solar-like oscillations. 
It was observed with the high-resolution spectrometer \harps\ at the ESO 3.6-m telescope in August 2006
(Mosser et al. 2008, hereafter M08). The authors analyzed the oscillation spectrum and identified 15 $\ell=0$ and $\ell=1$ eigenmodes. They found a model that agrees
with the physical parameters and all the seismic parameters but one: the behavior of the small spacing $\zeroun=\nu_{n,0}-(\nu_{n,1}+\nu_{n-1,1})/2$ with frequency. 
Since \cible\ is a low-mass star (less than 1 $M_\odot$), the effect of core overshooting was neglected in the preliminary modeling performed in M08.
For this range of mass, stars are not expected
to have a convective core on the main sequence, except for a small one which disappears shortly after the ZAMS.
In Sect. \ref{sect_models}, we show that in the case of \cible, with a reasonable amount of
mixing, the early convective core can survive until the present age. The agreement between observations and models is then greatly improved. In Sect. \ref{sect_discussion},
we explain why the convective core of \cible\ can survive, even when the burning of $^3$He is no longer capable of sustaining it.

%__________________________________________________________________
\section{Modeling of \cible \label{sect_models}}

\subsection{Results from previous modeling of \cible}

For their modeling of \cible, M08 adopted the following stellar parameters: $T\ind{eff}=6070\pm100\,K$ (\cite{2005A&A...440..321J}, \cite{2005A&A...441.1149D}),
$L=1.39\pm0.13\,L_\odot$ (from the Hipparcos parallax $\pi=107.98\pm0.19$~mas, \cite{2007ASSL..250.....V}) and an observed surface metallicity of $[Z/X]\ind{f}=-0.5\pm0.1$ dex, 
from detailed analysis (see M08). 
They searched for an optimal model by adjusting the age of the star, its mass, the initial abundance of helium $Y\ind{i}$, and the initial metallicity [Z/X]$\ind{i}$
to fit the following observational constraints: $T\ind{eff}$, $L$, $[Z/X]\ind{f}$, $\overline{\Delta\nu}$, $\zeroun$, and $\zerodeux$,
where $\overline{\Delta\nu}$ represents a mean value of the large spacing, and $\zeroun$, $\zerodeux$ are the small spacings defined as
\begin{eqnarray}
\delta\nu_{01,n}     & = & \nu_{n,0}-\frac{\nu_{n,1}+\nu_{n-1,1}}{2} \\
\delta\nu_{02,n}  & = & \nu_{n,0}-\nu_{n-1,2}.
\label{exp_spac}
\end{eqnarray}
Their best model fits the observational constraints on the star quite well, except for $\zeroun$, for which significant disagreements remain (see Fig. \ref{fit_d01}).

\subsection{A new modeling including extra mixing at the core boundary}

As shown in Appendix \ref{app_d01}, the small spacing $\zeroun$ is very sensitive to 
the central part of the stellar interior. The present disagreement suggests that the way we model these inner regions should be reconsidered.
As already mentioned, the models computed in M08 have no convective core. We tried to obtain models of \cible\ with a convective core here
by generating extra mixing at the edge of the core, as suggested by \cite{1985SoPh..100...21R}.

We performed a new modeling of \cible\ including mixing beyond the boundary of the convective core
by allowing $d\ind{ov}>0$ in our models. We computed stellar models using CESAM2k (\cite{1997A&AS..124..597M}), and we derived the mode frequencies from these
models with the Liege oscillation code (LOSC, \cite{2008Ap&SS.316..149S}). We used the same physics as in M08 for our models, apart from the treatment of convection.
We preferred to use the more realistic formalism of \cite{1991ApJ...370..295C}. 
For calibration, we computed a solar model with this treatment of convection and found $\alpha\ind{CM}=0.94$.
As in M08, we adopted this value for our modeling.
Overshooting is described, as explained before, as an extension of the adiabatic and mixed region associated to the convective core, over a distance $d\ind{ov}$ defined as
$$ d\ind{ov}=\min(r\ind{c},\,\aov H_P) $$
where $r\ind{c}$ is the radius of the convective core, $H_P$ the pressure scale height, and $\aov$ the overshooting parameter. The models computed with 
overshooting showed that the star can still have a convective core at its present age, provided the parameter $\aov$ is large enough.

%It is known that the existence of a convective core
%generates a discontinuity in the sound speed profile (as can be seen in Fig. \ref{sound_speed}), and induces an oscillation of the mode frequencies in function of the 
%radial order (see \cite{1990LNP...367..283G}, \cite{1993A&A...274..595P}). We therefore expect the behavior of the small spacing $\zeroun$ to be modified by the presence 
%of a convective core (see Section \ref{sect_theory}).

\subsection{Optimization\label{optimization}}

\begin{table}
\caption{Physical and seismic parameters of \cible\ derived from the observations, from the best model without overshooting (model \textbf{A}), and from
the best model with overshooting (model \textbf{B}).
\label{fit}}
\begin{tabular}{lccc}
\hline
  & Observations & model \textbf{A} & model \textbf{B} \\
\hline \hline
\multicolumn{4}{l}{Observational constraints} \\
\hline
$T\ind{eff}$ (K)  & 6070 $\pm$ 100 & 6095 & 6068 \\
$L/L_\odot$   & 1.39 $\pm$ 0.13 & 1.379 & 1.372 \\
$[Z/X]\ind{f}$ (dex) & $-0.5 \pm 0.1$ & $-0.560$ & $-0.548$ \\
\hline
$\overline{\Delta\nu}\,(\mu$Hz) & 120.3 $\pm$ 0.5 & 120.35 & 120.25 \\
$a_1 \, (.10^{-3})$ & $-5.6 \pm 2.8 $ & \mbox{\boldmath$ -2.4 $}$^{\star}$  & $-3.6 $  \\
$b_1\,(\mu$Hz) & $0.96 \pm 0.67$ & \mbox{\boldmath $2.85$}$^{\star}$ & 0.84 \\
$a_2 \, (.10^{-3})$ & $-3.8 \pm 4.2 $ & $-3.1$ & $-3.2 $ \\
$b_2\,(\mu$Hz) & $6.36 \pm 0.93$ & 6.43 & 5.97 \\
\hline \hline
\multicolumn{4}{l}{Free parameters} \\
\hline
$Y\ind{i}$ & - & $0.25 \pm 0.01$ & $0.26 \pm 0.01$ \\
$[Z/X]\ind{i}$ (dex) & - & $-0.42 \pm 0.05$ & $-0.50 \pm$ 0.07 \\
$M/M_\odot$   & - & 0.95 $\pm$ 0.11 & 0.94 $\pm$ 0.09 \\
age (Gyr) & - & 6.71 $\pm$ 0.75 & 6.72 $\pm$ 0.59 \\
$\aov$ & - & $\aov=0^\sharp$ & 0.17 $\pm$ 0.03 \\
\hline \hline
$\chi^2$ & - & 9.1 & 0.8 \\
\hline
\end{tabular}
\\
$^{\star}$ quantities not fitted within 1-$\sigma$ error bars \\
$^\sharp$ parameter fixed in the optimization
\end{table}

We looked for an optimal model fitting the global parameters of \cible:
$T\ind{eff}$, $\log(L/L_\odot)$, and $[Z/X]_{\hbox{\rm\small f}}$ (given in Table~\ref{fit}), as well as its seismic parameters. 
For the seismic constraints, we adopted the mode frequencies obtained in the analysis of M08.
We used a mean value of the large separation $\overline{\Delta\nu}$ and
the coefficients of a least-squares linear regression of the small spacings $\zeroun$ and $\zerodeux$. The frequency range of the identified modes is indeed small enough to linearize them in the form:
\begin{eqnarray}
\zeroun     & \simeq & a_1(\nu-\nu_0)+b_1 \\
\zerodeux & \simeq & a_2(\nu-\nu_0)+b_2,
\label{lin_spac}
\end{eqnarray}
with $\nu_0$ the frequency of the maximum of the signal ($\nu_0\simeq 2.75$ mHz). The observational values of these coefficients are given in Table \ref{fit}. We thus obtain a set of $N=8$ observational
constraints referred to as $p_i^{\hbox{\rm\tiny obs}}$, $i=1,N$. The free parameters are the stellar mass, age, initial helium abundance, metallicity, and the overshooting coefficient $\aov$ (see Table \ref{fit}).

The best models are obtained by minimizing the $\chi^2$ function defined as
\begin{equation}
\chi^2=\sum_{i=1}^N \left({p_i^{\hbox{\rm\tiny obs}}-p_i^{\hbox{\rm\tiny mod}} \over \sigma_i^{\hbox{\rm\tiny obs}}}\right)^2
\label{chi2}
\end{equation}
where $p_i^{\hbox{\rm\tiny mod}}$, $i=1,N$ are the values obtained in the models for the observationally constrained parameters. In contrast to M08, we used the 
Levenberg-Marquardt algorithm (as described in \cite{2005A&A...441..615M}) to find the minimum of the $\chi^2$ function, instead of computing a grid of models. 
This method is an interpolation between the Newton-Raphson algorithm and the steepest descent method. 
The steepest descent method is first used, ensuring a rapid approach when the minimum is far.
When getting closer to the minimum, the algorithm progressively switches to the Newton-Raphson method for a faster convergence toward the minimum.
The advantage is that one can find a minimum with only a few iterations. This kind of optimization is therefore less time-consuming than computing a 
grid of models with 5 free parameters and a fine grid mesh. The uncertainties on the parameters are obtained from the covariance matrix of the standard errors in the
free parameters.
We took the best model found in M08 as a first guess and performed two minimizations:
one without overshooting, giving best model \textbf{A}, and one with overshooting, giving best model \textbf{B}. Model \textbf{A} is, as expected, very close to the one found by M08. The only difference
lies in the error bars, which are bigger here. 
When using the grid-search method, the error bars are obtained by finding the change in each parameter, which increases $\chi^2\ind{min}$ by 1. This approach is
only correct if we can neglect the correlation between the different parameters, as explained in \cite{2003drea.book.....B}. 
Several studies have shown that correlations exist between the parameters (\cite{1994ApJ...427.1013B}, \cite{Ozel2009}), which are taken into account in the 
Levenberg-Marquardt optimization, since we have access to the non-diagonal terms in the covariance matrix. This explains why the uncertainties were underestimated in M08.

%This underlines the propensity of the "grid of models" method to underestimate the error bars, since it overlooks some models satisfying the observational constraints.

%This algorithm assumes that the $\chi^2$ function can be locally linearized to assess the error bars for each free parameter. The differences between the
%two methods suggests that there are parameters for which the non-linearity is not negligible in the $\Delta\chi^2<1$ space.

\begin{figure}
\centering
\includegraphics[width=8.5cm]{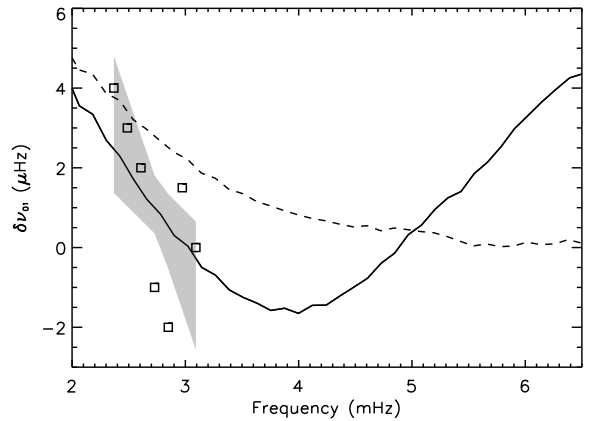}
\caption{Variations in the small spacing $\zeroun$ with frequency. The squares represent the observational values of $\zeroun$. The darkened area corresponds to
the linear regression of $\zeroun$ derived from the observations within 1-$\sigma$ error bars. The dashed line represents the
best model without overshooting (model \textbf{A}), and the full line the best model with overshooting (model \textbf{B}). The frequency axis has been extended to 6.5 mHz in order to exhibit the 
oscillation of $\zeroun$ for model \textbf{B}.
\label{fit_d01}}
\end{figure}

Table \ref{fit} gives the physical and seismic parameters of both models. For model \textbf{A}, parameters $a_1$ and $b_1$ are obtained 
at 1.2 $\sigma$ and 2.8 $\sigma$ of the observed values, respectively, causing a high value of $\chi^2\ind{min}\simeq 9.1$  (see Fig. \ref{fit_d01}). 
In contrast, for model \textbf{B}, $a_1$ and $b_1$ are fitted within 0.8 $\sigma$ and 0.2 $\sigma$, respectively. This results in a significant decrease of $\chi^2\ind{min}\simeq0.8$ for model \textbf{B}.

%The overshooting coefficient $\aov$ is very well constrained for this star, since it is determined at $\aov=0.171 \pm 0.026$.

%In the best model, the convective core survives, and it extends over about $8\%$ of the radius. 
%The presence of a convective core allows us to account for the frequency pattern of \cible, and the agreement when adding overshooting is much better. This leads to the 
%question of the influence of the core on the mode frequencies.

\section{Discussion of the results \label{sect_discussion}}

\subsection{Results of the fit}

\begin{figure}
\centering
\includegraphics[width=8.5cm]{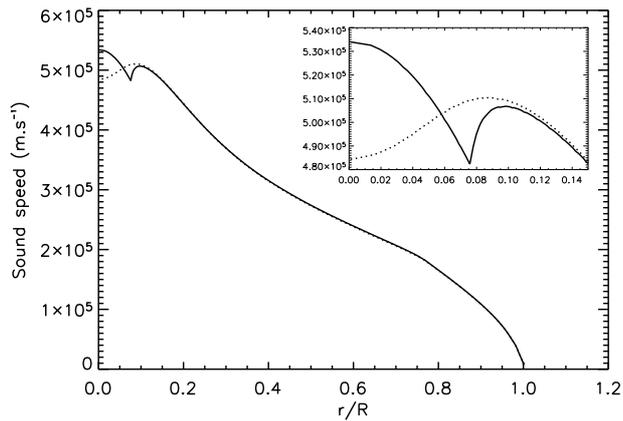}
\caption{Sound speed profile for the best model without overshooting (model \textbf{A}, dotted line), and the one with overshooting (model \textbf{B}, full line). A zoom of the curves shows the discontinuity at 
the edge of the convective core for model \textbf{B}.
\label{sound_speed}}
\end{figure}

When considering an extension of the mixed zone associated to the early convective core induced by overshooting, we get a model that fits all the observational constraints
better than within 1-$\sigma$ of the observed values. 
This decrease in the $\chi^2$ value in fact stems from the survival of the convective core. Indeed, model \textbf{B} has a convective core that extends over about $3\%$ of the stellar
mass. The withdrawal of this core generates a discontinuity in the chemical composition gradient, hence in the sound speed gradient (see Fig. \ref{sound_speed}). It has already been established 
that such a discontinuity induces an oscillation of the mode frequencies as a function of the radial order (see \cite{1990LNP...367..283G}).
\cite{1993A&A...274..595P} derived the expressions of mode frequencies in the case of a discontinuous  sound speed profile near the center, in the asymptotic
approximation. Using the second-order development they propose, we obtained (see Appendix \ref{app_d01}) the following expression for $\zeroun$:
\begin{equation}
\zeroun(\nu)=\frac{A}{\nu}-B\sin \left[ 2\pi\frac{n'}{\mathcal{P}}+\varphi(\nu) \right]
\end{equation}
where $n'$, $A$, $B$, and $\varphi(\nu)$ are defined in Appendix \ref{app_d01}. With a discontinuous sound speed profile, 
the small spacing $\zeroun$ oscillates with a period $\mathcal{P}$ corresponding to the ratio between the acoustic radius of the whole star and that of the discontinuity:
\begin{equation}
\mathcal{P}=\frac{\displaystyle\int_0^R \hbox{d}r/c}{\displaystyle\int_0^{r\ind{disc}} \hbox{d}r/c}
\label{eq_period}
\end{equation}
where $r\ind{disc}$ is the radius of the discontinuity.
We can see in Fig. \ref{fit_d01} that $\zeroun$ indeed oscillates for model \textbf{B}, which was not the case for model \textbf{A}.

When the amount of core overshooting increases, the acoustic radius of the discontinuity in the chemical
composition gradient increases and the period of the oscillation decreases (see Eq. \ref{eq_period}). Therefore, the variations in
$\zeroun$ become steeper. This shows that the behavior of $\zeroun$ in the models that have a convective core can be
adjusted by modifying the extension of the core, \textit{e.g.} by changing the overshooting coefficient. With an appropriate value of $\aov$, we can correctly fit the behavior
of $\zeroun$ and therefore reduce the $\chi^2$ value.
We find $\aov=0.17\pm0.03$ for \cible.
This result provides new input for the studies that seek to establish how convective core overshooting depends on
the stellar mass (\textit{e.g.} \cite{2007A&A...475.1019C}).

Stars in a range of mass half way between the presence and the absence of a convective core, such as \cible, are interesting objects for studying core overshooting.
Indeed, if the seismic analysis of such a star can establish the presence of a convective core, we obtain a firm lower bound for the amount of mixing at the edge of the core.
In this case, the extension of the core can be derived from seismic parameters, such as $\zeroun$, and we get a precise estimate of $d\ind{ov}$. And if, on the contrary,
no convective core is found, a strong upper limit for the extent of the region of extra mixing can be set.

\subsection{Survival of the core\label{sect_survival}}

The convective core of \cible\ disappears at about 200 Myr without overshooting, and survives until about 7 Gyr with $\aov=0.17$. We explain here this huge 
difference of longevity caused by overshooting. 

\subsubsection{Initial core}

\begin{figure}
\centering
\includegraphics[width=8.5cm]{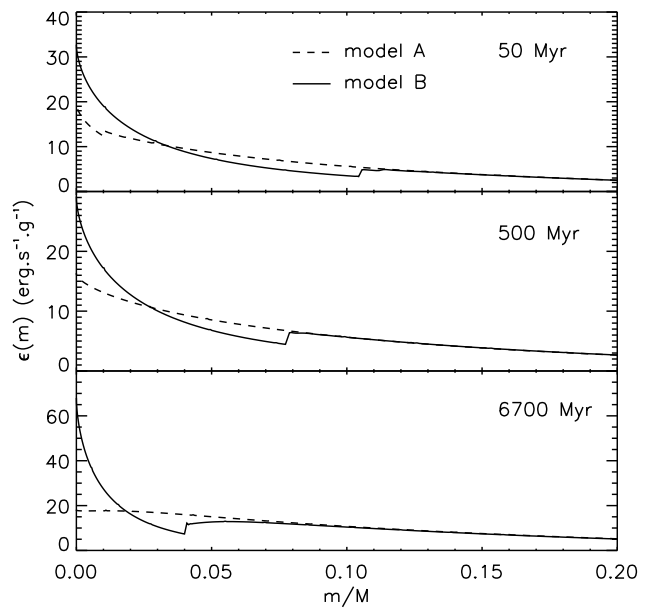}
\caption{Profile of the total energy production rate $\varepsilon(m)$ in the center of the star at different stages of its evolution (top: 50 Myr, middle: 500 Myr, bottom: 6700 Myr),
for model \textbf{A} (dashed line) and \textbf{B} (full line).
\label{fig_epsilon}}
\end{figure}

A convective core exists in main sequence models when the energy per unit mass $\varepsilon(m)$ generated by nuclear burning in the center is too high to be transported radiatively.
When the nuclear flux is larger than the maximum flux $F\ind{crit}$ that can be transported radiatively (corresponding to the radiative flux in the case of critical stability for the Schwarzschild criterion,
 i.e. $\nabla=\nabla\ind{ad}$), convective motions transport part of the energy.
 To ensure the existence of a convective core, we must have high values of the flux of energy $F(m)$ at low $m$, \textit{i.e.} a high value of $\varepsilon$ in the center.
 This condition is equivalent to having a steep (negative) gradient $\hbox{d}\varepsilon/\hbox{d}m$ in the center. 
Indeed, the star luminosity is almost entirely produced in the most central parts. If it is generated
with a gentle gradient of $\varepsilon$, the value of $\varepsilon$ in the center will be moderate (which is the case for model \textbf{A}, see Fig. \ref{fig_epsilon}). Only with a steep gradient of $\varepsilon$ 
can it reach higher values (see model \textbf{B} in Fig. \ref{fig_epsilon}). We therefore use the gradient of $\varepsilon$ hereafter as an indicator of the existence of a convective core.
 
In most cases, we may assume that the energy production is dominated by a given nuclear reaction chain. Thus, $\varepsilon(m)$ is proportional to $\rho X W T^\nu$, where $\rho$ is the density, 
$X$ the hydrogen mass fraction, $T$ the temperature, $\nu$ characterizes the temperature dependence of this specific reaction chain
and W is the mass fraction of the input element corresponding to the reaction (see \textit{e.g.} \cite{1994sipp.book.....H}, chap. 6).
We can then write
\begin{equation}
\gradeps \equiv  \frac{\hbox{d}\ln\varepsilon}{\hbox{d}\ln m} =  \frac{\hbox{d}\ln\rho}{\hbox{d}\ln m}+\frac{\hbox{d}\ln X}{\hbox{d}\ln m}+\frac{\hbox{d}\ln W}{\hbox{d}\ln m}+\nu \frac{\hbox{d}\ln T}{\hbox{d}\ln m}.
\label{eq_gradeps}
\end{equation}
We then see that, to have a steep $\gradeps$, the dominant nuclear reactions must have high temperature sensitivity.
 
For massive and intermediate-mass stars, the gradient of $\varepsilon$ is steep due to the strong temperature dependence of the CNO cycle ($\nu\simeq 20$, \cite{1994sipp.book.....H}).
For \cible, when reaching the ZAMS, the temperature in the center is of about $T_6\simeq13$ (where $T_6=T/10^6\,\hbox{K}$) in our models. 
At that temperature, the dominant reactions are those of the ppI chain.
Since the pp reaction -- $^1$H(p,e$^+\nu$)$^2$H -- is the slowest, its generation rate is proportional to that of the whole ppI chain if it has achieved equilibrium. The temperature sensitivity of the ppI
chain at equilibrium is therefore the same as that of the pp reaction $\nu\ind{pp}$. Based on \cite{1988ADNDT..40..283C}, we estimate $\nu\ind{pp}=4.1$ at our temperature. 
This is too weak for the star to have a convective core. However, as pointed out by \cite{1985SoPh..100...21R}, the abundances of $^3$He and $^{12}$C are in excess
compared to their equilibrium values when the star enters the main sequence. The temperature sensitivities of the
burning of $^3$He to $^4$He and of $^{12}$C to $^{14}$N outside of equilibrium are estimated to be $\nu\ind{He3}=16.7$ and $\nu\ind{C12}=18.7$
(Caughlan \& Fowler 1988). We therefore obtain a $\gradeps$ about four times steeper than for the equilibrated ppI chain. This ensures that an initial convective core is present.

\subsubsection{Withdrawal of the core}

To achieve equilibrium, the ppI chain requires that the ratio [$^3$He]/[H] corresponds to its equilibrium value. This ratio can be computed by assuming that the deuterium is constantly
in equilibrium, which is a good approximation since the destruction of D is much faster than the other reactions in the chain (see \cite{1968psen.book.....C}). 
It is inversely proportional to the temperature. In the core, the temperature increases towards the center, 
and the abundance of $^3$He should decrease with $r$ to achieve equilibrium. Since the elements are mixed in the convective core,
the abundance of $^3$He is constant with $r$, and the reactions are kept outside of equilibrium. 
The temperature sensitivity therefore remains high, which favors convection. One can say that convection is self-sustained here.

However, $\varepsilon$ also depends on the abundance of the reactant. As the star evolves, the abundance of $^3$He in the core decreases since it is destroyed faster than it is created.
Therefore, $\varepsilon\ind{He3}$ decreases, until the flux of nuclear energy becomes lower than $F\ind{crit}$ and the convective core disappears. This happens at an age of about
200 Myr for model \textbf{A} (without overshooting). When the core disappears, the ppI chain quickly achieves equilibrium (see Fig. \ref{fig_he3}).
The temperature sensitivity becomes that of the ppI chain, which causes $\gradeps$ to be less steep. Besides, the elements are no longer mixed
in the center, and in Eq. \ref{eq_gradeps}, $\hbox{d}\ln X/\hbox{d}\ln m>0$ and $\hbox{d}\ln W/\hbox{d}\ln m>0$. This also contributes to producing
a more gentle $\gradeps$.

\subsubsection{Effect of an extension of the mixed zone associated to the core}

The existence of a mixed zone at the boundary of the convective core modifies the abundances of elements in the center, and it was already suggested
by \cite{1985SoPh..100...21R} that it should increase the lifetime of the core.

\begin{figure}
\centering
\includegraphics[width=8.5cm]{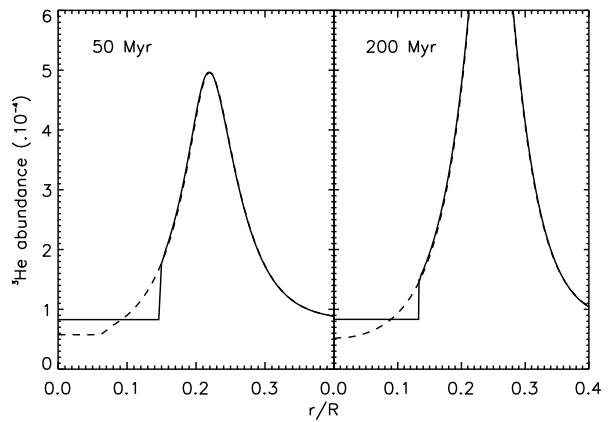}
\caption{Abundance of $^3$He in the center for two ages: 50 Myr (left) and 200 Myr (right). The dashed line stands for model \textbf{A} (without overshooting) and the
solid line for model \textbf{B} (with overshooting).
\label{fig_he3}}
\end{figure}

The peak we observe in the profile of the $^3$He abundance (see Fig. \ref{fig_he3}) is formed 
when the star reaches the ZAMS. It can be shown that the time required to achieve equilibrium is inversely proportional
to the temperature (see \cite{1968psen.book.....C}). To the left of the peak and outside the convective core, the ppI chain is already in equilibrium,
and the abundance of $^3$He decreases towards the center because the temperature increases. 
To the right of the peak, the reactions are not in equilibrium yet, and the abundance of $^3$He decreases towards the surface since
the nuclear reactions get less efficient because of decreasing temperature. 

When adding a mixed zone at the edge of the core, we can see in Fig. \ref{fig_he3} that the abundance of $^3$He in the core increases, owing to the
peak we just mentioned. Consequently,
$\varepsilon\ind{He3}$ increases. The convective core is therefore bigger and survives longer than in the case without additional mixing. 
If the peak described above did not exist, the mixing would not change the abundance of $^3$He and would have little effect on the core lifetime.
No such peak exists for the $^{12}$C profile, and its abundance is small because of the low metallicity of \cible. The role of $^{12}$C is negligible compared to that
of $^3$He in sustaining the core in the case of \cible.

When adding overshooting to our models, we indeed observe an extension of the core's lifetime. However, this extension is quite short. For example, with $\alpha\ind{ov}=0.1$, the core
disappears at an age of about 1 Gyr. In model \textbf{B}, the core is still present at an age of about 7 Gyr with $\alpha\ind{ov}\simeq0.17$. This sudden increase of the
core lifetime for $\aov>0.1$ is in fact caused by the start of the ppII and ppIII reaction chains, and later by the CNO chain, because of the increase in temperature in the center
as the star evolves. If the $^3$He has kept the convective core going until the ppII chain begins to compete, the convection prevents the ppII reactions from achieving equilibrium,
in exactly the same way as described above for the ppI chain.
The star starts burning $^7$Li through the reaction $^7$Li(p,$\alpha$)$^4$He out of equilibrium, with a temperature sensitivity of $\nu\ind{Li7}=10.8$, which sustains the core.
This is what happens for model \textbf{B}. If, on the contrary, the convective core has already disappeared at that time, there is no more mixing in the center
and the ppII reactions achieve equilibrium without triggering convection. This is the case for model \textbf{A}.

\begin{figure}
\centering
\includegraphics[width=8.5cm]{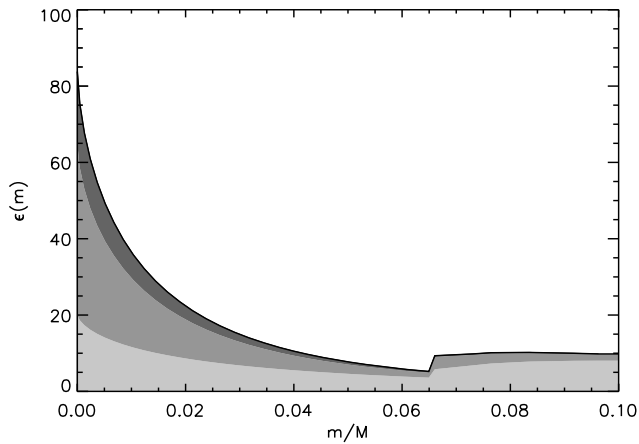}
\caption{Energy production rate $\varepsilon(m)$ in the central regions of the optimal model \textbf{B}. The dark line represents the total energy production rate.
The grey areas correspond to the luminosity produced by the different nuclear reaction chains (light grey: ppI chain, intermediate grey: ppII chain and dark grey: ppIII chain+CNO cycle).
\label{fig_contr_eps}}
\end{figure}

We can see in Fig. \ref{fig_contr_eps} that the ppII chain makes a major contribution to the nuclear production rate $\varepsilon(m)$. The importance of the CNO reactions is still small.
As mentioned above, the energy brought by the ppII chain is almost entirely due to the burning of $^7$Li outside equilibrium, with the other reactions contributing very little to $\varepsilon\ind{ppII}$.
If the star were only slightly more evolved, the CNO cycle would take over, causing the convective core to grow.

%\begin{table}
%\caption{Contribution of the burning of different elements in the value of the total energy production rate $\varepsilon$ (in erg.s$^{-1}$.g$^{-1}$) at the center of the star ($r=0$).
%Only the main contributors were detailed.
%\label{tab_epsilon}}
%\begin{tabular}{|l|c|c|c|c|c|c|}
%\hline
% Age & \multicolumn{2}{|c|}{50 Myr} & \multicolumn{2}{|c|}{500 Myr} & \multicolumn{2}{|c|}{6700 Myr} \\
%\hline
% Model & A & B & A & B & A & B \\
%\hline
%$\varepsilon\ind{H1}$ ($pp$) & 8.91 & 9.44 & 8.77 & 9.75 & 9.84 & 45.03 \\
%$\varepsilon\ind{He3}$ ($pp1$) & 9.96 & 21.15 & 7.32 & 18.94 & 0.16 & 4.32 \\
%%$\varepsilon\ind{other}$ & 0.6 & 1.8 & 0.1 & 0.2 & 7.9 & 17.8 \\
%$\varepsilon\ind{CNO}$ & 0.56 & 1.69 & 0.04 & 0.04 & 7.23 & 14.31 \\
%Total & 19.51 & 32.40 & 16.20 & 28.87 & 17.90 & 67.20 \\
%\hline
%\end{tabular}
%\end{table}

%\begin{figure}
%\centering
%\includegraphics[width=8.5cm]{fig_flux2.ps}
%\caption{Difference between the nuclear flux  and the critical flux (maximum flux which can be transported radiatively) in the center of the star. A positive value indicates the necessity of convective
%transport. This difference is represented for the nuclear flux generated only by hydrogen burning ($F\ind{H1}-F\ind{crit}$ in dotted line), for the flux generated by the burning of $^1$H and $^3$He
%($F\ind{H1}+F\ind{He3}-F\ind{crit}$ in dashed line) and for the total nuclear flux ($F\ind{tot}-F\ind{crit}$ in full line). A zoom of the most central parts of the star is provided for the early stages of evolution of model \textbf{A}.
%\label{fig_flux}}
%\end{figure}

\subsection{Influence on the evolution}

We notice that, even though the structure of the core is different for our models with and without overshooting, their age is very comparable (see Table \ref{fit}). 
This suggests that the survival of the convective core does
not significantly influence the global nuclear energy produced by \cible\ during its evolution. 
This constitutes a striking difference with higher mass models, where overshooting is known to have 
a strong impact on the evolution of the star, especially on its age, for a given $T\ind{eff}$ and $L$.
For high-mass stars, since the temperature dependence of the dominant CNO cycle is large, it operates in a narrow area in the center, and the wider convective core can act as a reservoir. For \cible, the pp chain
is dominant, and its temperature dependence is much lower. Therefore, the reactions take place in an area wider than the extent of the convective core, and the mixing has less effect on the evolution.

This means that it would be hopeless to try to characterize an extension of the convective core in this type of star by classical fundamental stellar parameters alone ($T\ind{eff}$, $L$, $M$), 
as was done by Claret (2007) for higher masses.

%__________________________________________________________________

%______________________________________________________________
\section{Conclusion\label{conclusion}}

We present here a modeling of \cible\ based on the analysis 
of \harps\ data performed in M08. Our main result is that, on this basis,
we find strong evidence that this old low-mass star has a convective core.
Models with convective cores enabled us to solve the disagreement with observations
that was pointed out in M08 for models without convective cores, bringing the $\chi^2$
function from $9.1$ to $0.8$. All the observed parameters for \cible\ are now fitted within 1-$\sigma$ error bars.

In the case of our modeling of \cible, the value obtained for the $\aov$ 
parameter ($0.17 \pm 0.03$) is strongly constrained.   
Overshooting was here used as a proxy to model the complex processes of transport
at the edge of a convective core, as is usually done in the present state of stellar modeling.
Rather than finding a unique absolute value for $\aov$,
the current aim is to try to observationally determine which values of the
$\aov$ parameter are needed to represent stars of different masses and evolution stages.
In this respect, the value obtained for \cible\ constitutes a valuable input for
low-mass objects.

We discussed in detail how the existence of a convective core in such an evolved
low-mass star can be explained by a reasonable amount of extra mixing (modeled here as core overshooting)
inducing the survival of the early convective core. For low-mass stars such as \cible, an early convective core exists
because of the burning of $^{12}$C and $^3$He outside equilibrium. An extra mixing at the edge of the core increases its lifetime, 
by bringing more $^3$He to the center, as mentioned in \cite{1985SoPh..100...21R}. Here, we showed that, above a certain amount of overshooting ($\aov\sim0.15$),
the burning of $^3$He out of equilibrium sustains the core until the ppII and ppIII reactions take over. Convection
prevents these reactions from achieving equilibrium, and the burning of $^7$Li outside equilibrium is currently keeping the core convective.
This is a transitional phase before the CNO cycle takes over.

The observation of low-mass stars can play a specific role in the study
of extra mixing at the edge of stellar cores. Indeed, the presence or absence of a convective core in these stars,
which can be established by seismic indicators such as the $\zeroun$ small spacing, strongly constrains
the amount of mixing at the edge of the core. This stresses the interest in observing this type of star with ground-based 
observation campaigns or with the space mission \corot\ (\cite{2006ESASP1306...33B}).

%______________________________________________________________
%\begin{acknowledgements}\end{acknowledgements}

\begin{acknowledgements}
We are grateful to A. Noels for useful discussions about the nuclear reactions in the core of low-mass stars.
\end{acknowledgements}

%The aim is here to confirm that the oscillation observed in the models is indeed caused by the presence of a convective core. 

\begin{appendix}

\section{Expression of the small spacing $\zeroun$ with a discontinuous sound speed profile\label{app_d01}}

By neglecting the variations in the gravitational potential (Cowling approximation) and using the asymptotic development derived in Tassoul (1980), 
\cite{1993A&A...274..595P} obtained the expression of the oscillation frequencies in the case of a discontinuous sound speed profile.
The discontinuity is characterized by the parameter $\eta$:
$$\eta=\frac{\rho_i c_i - \rho_o c_o}{\rho_i c_i + \rho_o c_o}$$
where the subscripts $i$ and $o$ describe the quantities at the edge of the convective core
in the inner and outer regions. \cite{1993A&A...274..595P} showed that under these assumptions and provided $\eta$ is small enough, 
the first-order asymptotic development of the frequencies is given by the expression
\begin{equation}
\nunl=\left[ n' +\frac{\ell}{2} -\frac{\eta}{\pi}\sin 2\pi \left( \frac{n' +\ell/2}{\mathcal{P}} - \frac{\ell}{2}\right) \right]\Delta\nu
\label{eq_first_order}
\end{equation}
with
\begin{eqnarray*}
n' & = & n+\epsilon   \\  
\Delta\nu & = & \left( 2\displaystyle\int^R_0 \frac{\hbox{d}r}{c} \right)^{-1}  
\end{eqnarray*}
where $n$ is the radial order of the mode and $\ell$ the degree. The parameter $\epsilon$ depends on the reflective properties of the surface, and its variations with
frequency are neglected. When there is a discontinuity, the eigenfrequencies present a sinusoidal oscillatory behavior. These oscillations have an amplitude
of $\eta\Delta\nu/\pi$ and a period $\mathcal{P}$, which corresponds to the ratio of the acoustic radius of the whole star to that of the discontinuity (see Eq. \ref{eq_period}).

Based on this, we can derive the following expression for $\zeroun$:
\begin{eqnarray}
\zeroun & = & -\frac{\eta\Delta\nu}{2\pi}  \left[ 2\sin \left( 2\pi\frac{n'}{\mathcal{P}} \right) + \sin \left( 2\pi \frac{n'}{\mathcal{P}} + \frac{\pi}{\mathcal{P}} \right) + \right. \\   \nonumber
               &   &                                                 \left. \sin \left( 2\pi \frac{n'}{\mathcal{P}} - \frac{\pi}{\mathcal{P}} \right)  \right]    \\    \nonumber
               & = & -\frac{2\eta\Delta\nu}{\pi} \cos^2 \left( \frac{\pi}{2\mathcal{P}} \right)  \sin \left( 2\pi\frac{n'}{\mathcal{P}} \right).
\label{eq_d01_first}
\end{eqnarray}
As can be seen in Fig. \ref{fit_d01}, the mean value of the small spacing $\zeroun$ is not zero, as could be expected from Eq. \ref{eq_d01_first}. To understand this,
one needs to push the asymptotic development of the frequencies to the second order. Still from \cite{1993A&A...274..595P}, the expression for the frequencies is given by
\begin{equation}
\nunl=\left[ n'+\frac{\ell}{2}+\frac{\ell(\ell+1)V_1+V_2}{4\pi^2\nunl} -\frac{\eta}{\pi}\sin\alpha_{n,\ell}\right]\Delta\nu
\label{eq_second_order}
\end{equation}
where
$$ \alpha_{n,\ell}=2\pi \left( \frac{n'+\ell/2}{\mathcal{P}}-\frac{\ell}{2}-\frac{\ell(\ell+1)V_3+V_4}{4\pi^2\nunl} \right). $$
The $V_j$ for $j=1,4$ are defined in \cite{1993A&A...274..595P}. They are of the same order of magnitude as $\int_0^R \hbox{d}c/r$.
The small spacing $\zeroun$ is therefore sensitive to rapid variations in the sound speed, and particularly in the most central regions, owing to the
$1/r$ factor in the integrand. Using the second-order development of the freqeuncies (Eq. \ref{eq_second_order}), the expression of $\zeroun$ is a bit more complicated. We have
\begin{eqnarray}
\zeroun & = &-\frac{V_1\Delta\nu}{2\pi^2\nu}-\frac{\eta\Delta\nu}{2\pi}  \left[  2\sin \left( 2\pi\frac{n'}{\mathcal{P}} - \frac{V_4}{2\pi\nu}\right) + \right. \\  \nonumber
               &    & \sin \left( 2\pi \frac{n'}{\mathcal{P}} + \frac{\pi}{\mathcal{P}} - \frac{2V_3+V_4}{2\pi\nu}\right) + \\   \nonumber
               &    &  \left. \sin \left( 2\pi \frac{n'}{\mathcal{P}} - \frac{\pi}{\mathcal{P}} - \frac{2V_3+V_4}{2\pi\nu}\right)  \right]    \\    \nonumber
               & = &-\frac{V_1\Delta\nu}{2\pi^2\nu} -\frac{2\eta\Delta\nu}{\pi} \biggl[  \\ \nonumber
               &     & \cos^2 \left( \frac{\pi}{2\mathcal{P}} \right) \cos \left( \frac{V_3}{2\pi\nu} \right)  \sin \left( 2\pi\frac{n'}{\mathcal{P}} -\frac{V_3+V_4}{2\pi\nu} \right) + \\ \nonumber
               &     & \left.  \sin^2 \left( \frac{\pi}{2\mathcal{P}} \right) \sin \left( \frac{V_3}{2\pi\nu} \right)  \cos \left( 2\pi\frac{n'}{\mathcal{P}} -\frac{V_3+V_4}{2\pi\nu} \right) \right].
\label{eq_d01_second}
\end{eqnarray}
This can be rearranged in the form
\begin{equation}
\zeroun(\nu)=\frac{A}{\nu}+B\sin \left[ 2\pi\frac{n'}{\mathcal{P}}+\varphi(\nu) \right]
\end{equation}
where
\begin{eqnarray*}
A & = & -\frac{V_1\Delta\nu}{2\pi^2} \\   %\nonumber
B & = & -\frac{2\eta\Delta\nu}{\pi} \left[ \cos^4 \left( \frac{\pi}{2\mathcal{P}} \right) \cos^2\left( \frac{V_3}{2\pi\nu} \right) + \right. \\
    &    & \left. \sin^4 \left( \frac{\pi}{2\mathcal{P}} \right) \sin^2 \left( \frac{V_3}{2\pi\nu} \right) \right]^{\frac{1}{2}} \\  % \nonumber
\varphi(\nu) & = & \arctan \left[ \tan^2 \left( \frac{\pi}{2\mathcal{P}} \right) \tan \left( \frac{V_3}{2\pi\nu} \right) \right] - \frac{V_3+V_4}{2\pi\nu}. \\   %\nonumber
\end{eqnarray*}

With a discontinuous sound speed profile, the small spacing $\zeroun$ oscillates with the same period $\mathcal{P}$ as the individual frequencies.
In the continuous case, $\eta=0$, and the expression of $\zeroun$ reduces to the usual second-order asymptotic development, which varies like $1/\nu$,
as is the case for model \textbf{A} (see Fig. \ref{fit_d01}).

\end{appendix}

\end{document}